\documentclass[prb,twocolumn,showpacs]{revtex4}
\usepackage{graphicx}
\usepackage{dcolumn}
\usepackage{bm}
\usepackage{psfrag}

\begin{document}

\title{On continuum modeling of sputter erosion under normal incidence:\\
 interplay between nonlocality and nonlinearity}

\author{Sebastian Vogel}
\author{Stefan J. Linz}%
\affiliation{%
Institut f\"ur Theoretische Physik, Universit\"at M\"unster,
Wilhelm-Klemm-Str.9, D-48149 M\"unster, Germany
}%

\date{\today}

\begin{abstract}
Under specific experimental circumstances, sputter erosion on semiconductor
materials exhibits highly ordered hexagonal dot-like nanostructures. In a 
recent attempt to theoretically understand this pattern forming process, 
Facsko et al. [Phys. Rev. B {\bf 69}, 153412 (2004)] suggested  a nonlocal, 
damped Kuramoto-Sivashinsky equation as a potential candidate for an 
adequate continuum model of this self-organizing process. 
In this study we theoretically investigate this proposal by (i) formally 
deriving such a nonlocal equation as minimal model from balance 
considerations, (ii) showing that it can be exactly mapped to a local, damped 
Kuramoto-Sivashinsky equation, and (iii) inspecting the consequences of the 
resulting non-stationary erosion dynamics.
\end{abstract}
\maketitle

\section{Introduction} 
Sputter erosion \cite{carter}, the bombardment of solid target surfaces 
with ionized particles to remove or to detach target material, has a long 
tradition in physics as an experimental technique to clean, smooth,  
or appropriately prepare solid surfaces. The preparation of on the nanoscale 
perfectly flat surfaces, however, does not seem to be possible. This is because 
various surface roughening and smoothing processes compete during 
the sputtering process and can lead to self-organized pattern formation of the
eroded surface morphology. Depending 
on the target material, temperature, ion beam energy,  
angle of incidence and various other parameters, one generically observes the
development of rough surfaces, cellular patterns  and, in particular under oblique incidence 
of the ion beam, the formation of more or less regular ripple patterns. For an
overview on recent experimental results and theoretical approaches using 
continuum modeling we refer to Makeev et al. \cite{makeev} and 
Valbusa et al. \cite{valbusa}.  

During the last five years, however, spectacularly novel experimental results 
have been reported by Facsko et al \cite{facsko1,facsko2,facsko3,facsko4} 
showing that 
GaSb and InSb semiconductor targets eroded by Ar$^+$ ions under normal 
incidence can develop into a rather well ordered surface morphology with 
basically hexagonally arranged dot structures.        
Similar results have been subsequently reported by Gago et al\cite{gago} 
for Si targets under normal incidence and, more generally, by Frost et al. 
\cite{frost1,frost2,frost3} for rotated InP, InSb and  GaSb targets under 
oblique incidence (where as function of the inclination angle a variety of 
other patterns have also been observed).
To theoretically explain the hexagonal ordering and taking advantage 
of earlier work by Elder et al. \cite{elder1,elder2} 
(cf. also an even earlier study\cite{siva1} on this subject),  
Facsko et al. \cite{facsko-t} have recently suggested 
a (stochastically extended) stabilized or damped Kuramoto-Sivashinsky 
(KS) equation for the dynamics of the surface height $H$. This equation 
is given by
\begin{equation}
\partial_t H= -v_0-\alpha H +\nu \nabla^2 H-D_{\rm eff}\nabla^4 H
+\frac{\lambda}{2}(\nabla H)^2+ \eta
\label{skse}
\end{equation}
and might be considered as a useful continuum model for the ion-beam erosion 
under normal incidence since it can successfully reproduce 
the experimentally observed hexagonally ordered dot-type structures. The 
physical origin of the six terms on the rhs of Eq.(\ref{skse}) is attributed 
to constant erosion velocity  $v_0$, dissipation, effective surface tension 
(Bradley-Harper mechanism \cite{bradley}), thermal and erosion induced diffusion, 
tilt dependent sputtering yield, and some stochasticity of the erosion process,
respectively (cf. the original study \cite{facsko-t} for details).    
Previous attempts\cite{facsko2,kahng} based on the (standard) KS equation, 
i.e. without the term
$-\alpha H$, have reproduced cellular patterns without convincing evidence 
of a regular hexagonal ordering. As noted by Facsko et al. at 
the end of their paper \cite{facsko-t}, there are two basic problems with
the applicability of Eq.(\ref{skse}) to erosion processes since, as 
Eq.(\ref{skse}) stands, it violates the translation invariance 
in the erosion direction and, moreover, the physical meaning of the dissipation 
term $-\alpha H$ is not directly obvious. Facsko et al. \cite{facsko-t} 
briefly argued that (i) translational invariance can be restored by assuming 
that $-\alpha H$ has to be replaced by $-\alpha (H-\overline{H})$ with 
$\overline{H}$ being the erosion depth averaged over the sample area and (ii) 
that the term 
$-\alpha (H-\overline{H})$ might be interpreted as an approximation of a newly 
suggested redeposition effect of the sputtered target particles. Due to the 
nonlocal character of this term, however, the systematic connection between 
this nonlocally extended damped KS equation and Eq.(\ref{skse}) is far from
being obvious and the main reason for our paper. 
  
First,  we argue on general grounds, by using balance considerations, 
symmetries and allowing for nonlocal dependencies, what the simplest functional 
form of the spatio-temporal dynamics of the morphology of ion sputtered 
surfaces under {\it normal} incidence of the ions might be, if 
nonlocal terms as suggested by Facsko et al. 
\cite{facsko-t} are taken into account. Second, we show that the resulting 
non-local model equation can be mapped to the damped KS equation by the 
use of a temporally nonlocal transformation. By that we show how the 
stabilized KS equation suggested by Facsko et al. \cite{facsko-t} 
systematically fits into the theoretical 
framework of a continuum description of eroded surfaces obeying all 
desired invariances.  Third, we discuss in detail the subtle interplay 
between potentially nonlocal terms in the evolution equation for the 
eroding surfaces and the role of the KS-nonlinearity $(\nabla H)^2$. 
Finally, we also investigate some consequences on the general evolution 
dynamics in these equations. 

\section{Balance equation considerations} 
The starting point of our investigation is the general, stochastically 
extended balance equation for the spatio-temporal evolution of the eroded 
surface morphology $H({\bf x},t)$  measured perpendicularly to the 
initially flat target surface with 
coordinates ${\bf x}=(x,y)$. Assuming that the particle density at 
the surface of the target material can be basically considered as being 
constant \cite{linz1,linz2}, such a balance equation reads quite 
generally \cite{bs}

\begin{equation}
\partial_t H={\nabla}\cdot{\bf J}_H + F +\eta.
\label{eq1}
\end{equation}
and  expresses the fact that temporal changes of the erosion depth 
$H({\bf x},t)$ arise from 
two main contributions: (i) the detachment of target material leading to an 
in general inhomogenous and non-stationary, appropriately rescaled  
flux of eroded particles $F$ (given by the number of eroded particles 
per time and surface area divided by the particle density) 
and (ii) the local rearrangements of the particles 
at the surface leading to a relaxational current ${\bf J}_H$ along the surface. 
Note that this balance only accounts for the target particles. The underlying 
driving of the erosion process, the flux of eroding ions $I$, enters indirectly 
into such a description: all terms on the right hand side of (\ref{eq1}) depend
on $I$ in a way that they vanish if the sputtering process is 
turned off, $I=0$. Note that thermally activated processes 
that do not depend on $I$ and might lead to a further smoothing of the 
surface after the sputtering has been stopped are not taken into account. 
The spatio-temporal fluctuations $\eta=\eta({\bf x},t)$ entering in (\ref{eq1}) 
mimic some stochasticity present in the erosion process and are usually 
assumed to be Gaussian white, i.e. having an average 
 $\langle\eta({\bf x},t)\rangle_\eta=0$ and a covariance
$\langle\eta({\bf x},t)\eta({\bf x^\prime},t^\prime)\rangle_\eta
=2 D \delta({\bf x}-{\bf x^\prime})\delta({t}-{t^\prime})$. 

Considering periodic boundary conditions on an appropriately chosen, large
enough sample area of the size $L^2$ and introducing the spatial average  
$\stackrel{-}{..}(t)=
 (1/L^2)\int_{-L/2}^{L/2}dx \int_{-L/2}^{L/2} dy ...$ (being generally not
 equivalent to a stochastic average if taken on a finite area),  
the evolution of the mean erosion depth $\overline H(t)$ develops according to 
 $\partial_t\overline H(t)=\overline F+\overline \eta $ which directly 
 leads to 
\begin{equation}
\overline H(t)-\overline H(0)=\int_0^t\overline F\; dt^\prime
   +\int_0^t\overline \eta\; dt^\prime  
\end{equation}
with $\overline H(0)$ being the spatial average of the initial condition 
$H({\bf x},0)$ (which is usually set to zero under the assumption of an  
initially flat target surface). Already at this stage it is clear 
that the mean erosion depth $\overline{H}(t)$ is, in general, not a linear
function in $t$ implying a constant erosion velocity, but a rather 
complicated function that integrates over the 
history of the stochasticity and of the eroded flux.\par

In order to specify the admissible functional forms of the right hand side 
of Eq.(\ref{eq1}), we reconsider the three fundamental symmetry requirements 
that are considered to be basic for the spatio-temporal evolution of surface 
morphologies \cite{bs}: (i) no dependence of (\ref{eq1}) on the specific choice of the 
origin of time implying invariance of (\ref{eq1}) under translation in time, 
(ii) no dependence of (\ref{eq1}) on the specific choice of the  
origin of the coordinate system implying invariance of (\ref{eq1}) 
under translation in the direction perpendicular to the erosion direction, and 
(iii) no dependence of (\ref{eq1}) on the specific choice of the origin of the 
$H$-axis implying invariance of (\ref{eq1}) under translation in growth
direction. These symmetry requirements exclude any {\it explicit} dependence of 
${\nabla}\cdot{\bf J}_H$ and $F$ on the time $t$, the spatial position 
${\bf x}$ and the erosion depth $H$, respectively. Following Facsko et al.'s
argument \cite{facsko-t}, however, 
an implicite functional dependence on $H-\overline H$
and on the spatial derivatives of $H$ is still admissable. Consequently, the
detachment contribution in (\ref{eq1}) is quite generally given by 
$F[\nabla H, H-\overline H]$ where $[\nabla H, ..]$ is the short hand
notation for any derivative or combination of derivatives of $H$ being 
compatible with the scalar character of $F$ or $H$.

To proceed we (i) additionally apply invariance under rotation and reflection 
in the plane perpendicular to the erosion direction (which is suggested 
by the experimentally observed amorphorization of the target surface due
to the erosion), (ii) expand the height dependent term $f_\beta$ of $F$ in 
a power series 
in $H-\overline H$, i.e. $f_\beta=\sum_n \beta_n (H-\overline H)^n$, 
and (iii) perform a gradient expansion of the contribution $f_\alpha$ of $F$ that  
solely contains derivatives of $H$. The latter implies that the lowest order 
terms (up to forth order in $\nabla$ and 
second order in $H$) are given by $\nabla^2 H$,
$(\nabla H)^2$, $\nabla^4 H$, $\nabla^2(\nabla^2 H) $, $(\nabla^2 H)^2$), 
and $\nabla \cdot [(\nabla H)(\nabla^2 H)]$. 
Keeping only the lowest order contributions, the erosion flux is determined by    
\begin{eqnarray}
F&=& F_0 \left[ 1 + \beta_1 (H-\overline H)+\alpha_1 \nabla^2 H+\alpha_2 
(\nabla H)^2\right]\nonumber\\
              &\;& +\, O\left(\nabla^4,H^2,(H-\overline H)^2\right)
\label{trafo1}
\end{eqnarray}
with $F_0$ being a constant and negative since this part of the flux is 
antiparallel to the erosion direction. The lowest order possible mixing term 
between $\nabla H$ and $H-\overline H$ that could appear in (\ref{trafo1})  
is given by $(H-\overline H)\nabla^2 H$ and will be omitted since it
can be considered as a higher order contribution to the term proportional   
to $\alpha_1$ or $\beta_1$ in (\ref{trafo1}).
 
The functional form of the term representing the relaxational currents 
at the surface, ${\nabla}\cdot{\bf J}_H$ in (\ref{eq1}) remains to be specified. 
Following Facsko et al. \cite{facsko-t}, 
we suppose ${\nabla}\cdot{\bf J}_H=-D_{\rm eff}\nabla^4 H$. 
This mimics the tendency of surface particles to reach energetically 
more favorable positions with a positive curvature $\nabla^2 H>0$ and,
therefore, leads to a current ${\bf J}_H\propto \nabla(\nabla^2 H)$.     

Combining all ingredients and renaming the entering coefficients 
$b=\beta_1 F_0$, $a_1=\alpha_1 F_0$, $a_2=-D_{\rm eff}$, and $a_3=\alpha_2 F_0$, 
yields a minimal functional form for the evolution of the erosion depth under 
the afore-mentioned restrictions that is given by
\begin{eqnarray}
\partial_t H&=& a_1 \nabla^2 H+a_2\nabla^4 H+a_3(\nabla H)^2\nonumber \\
            &\,& + b (H-\overline H)+F_0+ \eta.
\label{trafo2}
\end{eqnarray}
The functional form of Eq.(\ref{trafo2}) looks like the equation 
suggested by Facsko et al\cite{facsko-t} that follows after their 
ad-hoc replacement of $-\alpha H$ by $- \alpha(H-\overline H)$ in the damped 
KS equation. Note however, that the spatially constant part of the flux $F$, 
i.e. $F_0$, is, in general, not given by the mean constant erosion velocity 
since the latter is not constant during the course of the evolution. 
   
Next we transform Eq.(\ref{trafo2}) into a coordinate system that moves 
with the mean erosion depth $\overline H(t)$,
\begin{equation}
h({\bf x},t)=H({\bf x},t)-\overline H(t)
\label{trafo3}
\end{equation}
with $h({\bf x},t)$ being the local erosion profile that obviously fulfills 
$\overline h=0$ and $\partial_t\overline h=0$ for all times $t$. The 
transformation (\ref{trafo3}) also guarantees that the 
invariance under translation in erosion direction $H\rightarrow H+z$ with $z$
being an arbitrary constant (that only needs to hold in the 
coordinate system that is fixed in space) holds for {\it any} erosion 
profile $h({\bf x},t)$. As a consequence, the evolution dynamics for 
$h({\bf x},t)$ does not necessarily 
need to fulfill this requirement.
 
Using Eq.(\ref{trafo2}) and the facts that 
$\partial_t H=\partial_t \overline H +\partial_t h$ 
and that any terms consisting only of spatial derivatives of $H$ 
have the same functional form in the comoving system when $H$ is 
substituted by $\overline H+h$,  the evolution equation for the erosion 
process in the comoving frame reads explicitly
\begin{equation}
\partial_t h= b h+a_1 \nabla^2 h+a_2\nabla^4 h+a_3(\nabla h)^2
+F_0-\partial_t\overline H+ \eta
\label{trafo4}
\end{equation}
Eq.(\ref{trafo4}) still constitutes a stochastic integro-partial differential  
equation by virtue of the nonlocal term $\partial_t\overline H$. Only if  
$\partial_t\overline H=F_0$ or, equivalently, the mean erosion depth 
$\overline H(t)$ were moving with a {\it constant} speed for all times, 
Eq.(\ref{trafo4}) would reduce to the damped stochastic KS equation. 
In general, however, this cannot be invoked, in particular because 
of the presence  of the KS-type nonlinearity $(\nabla h)^2$.
By taking the spatial average of Eq.(\ref{trafo4}), $a_3 \overline{(\nabla
h)^2}+F_0-\partial_t\overline H+\overline\eta=0$,  one can directly connect 
the  mean erosion depth $\overline H(t)$ to the dynamics of the local 
erosion profile $h({\bf x},t)$,   
\begin{equation}
\overline H(t)= F_0 t + \int_0^t dt^\prime \,\left[a_3\overline{(\nabla h)^2}
+\overline\eta\right]
\label{trafo5}
\end{equation}
where $\overline H(0)=0$ has been assumed. Consider first the deterministic 
part of (\ref{trafo5}), i.e. with $\overline\eta=0$.
Since the integrand of the second term on the right hand side of 
Eq.(\ref{trafo5}) is positive for all times, the mean erosion depth 
$\overline H(t)$ systematically deviates from a linear time 
evolution given by $F_0 t$. For $a_3$ being negative (positive) the 
mean erosion depth is therefore retarded (advanced) in comparison to $F_0 t$. 
Noteworthy, this fact is not a specific property triggered by the 
nonlocal term in Eq.(\ref{trafo2}), it is already present in the 
standard KS equation and from related studies there\cite{linz1}, 
it is known that the second term of the right hand side of (\ref{trafo5}) 
significantly contributes to the time evolution of $\overline H(t)$. 
The additional impact of the term $\int_0^t dt^\prime\overline\eta$ is roughly 
that of a superposed Wiener process since $\eta$ has been assumed to be a 
Gaussian and white. 
\section{Transformation to a local equation} 

As explained in the previous section, it is not obvious how the nonlocal 
equations (\ref{trafo2}) and (\ref{trafo4}) are related to the damped 
KS equation. Here we show in a more general context how the nonlocal term 
can be elimated by an appropriate transformation. 
Specifically our statement is as follows:  There is a transformation 
$h\rightarrow \hat h$ that maps the general form of a 
{\it nonlocal} stochastic evolution equation given by
\begin{equation}
\partial_t h=b h + G[\nabla h]+ F_0- \partial_t \overline H +\eta
\label{eqta}
\end{equation}
to a {\it local} evolution equation in the transformed 
variables $\hat h$ reading 
\begin{equation}
\partial_t \hat h=b \hat h + G[\nabla \hat h]+\eta
\label{eqtb}
\end{equation}
where, quite generally, the functional $G$ in the Eqs.(\ref{eqta}) 
and (\ref{eqtb}) can contain any combination of derivatives of $h$ or 
$\hat h$, respectively, but no explicit dependence on $h$ or $\hat h$.
 
As subsequently useful observation, we note that Eq.(\ref{eqtb}) is 
invariant under the transformation $\hat h\rightarrow \hat h+z\,\exp(bt)$ 
with $z$ being an arbitrary constant.   Obviously, (\ref{eqta}) 
contains the model equation (\ref{trafo4}) as special case if we set 
$G[\nabla h]=a_1 \nabla^2 h+a_2\nabla^4 h+a_3(\nabla h)^2$.
 
To the constructive derivation of the statement:  
In order to find the desired transformation we use the ansatz  
\begin{equation}
h\ \rightarrow \hat h=h + \frac{F_0}{b} +f(t)
\label{eqtc}
\end{equation}
where the time-dependent function $f(t)$ is so far arbitrary and will be  
subsequently determined.  
Since $\partial_t h= \partial_t \hat h-\partial_t f$ and spatial derivatives 
of $h$ transform without any change to spatial derivatives in $\hat h$, 
 insertion of 
(\ref{eqtc}) into Eq.(\ref{eqta}) yields    
\begin{equation}
\partial_t \hat h=b \hat h + G[\nabla \hat h]+\eta +\partial_t f-bf-
\partial_t \overline H.
\label{eqtd}
\end{equation}
In order to arrive from Eq.(\ref{eqta}) at Eq.(\ref{eqtb}), one has to demand
that the last three terms on the right hand side of Eq.(\ref{eqtd}) vanish at
any time,  
\begin{equation}
\partial_t f=bf+\partial_t \overline H.
\label{eqte}
\end{equation}
The linear nonautonomous ordinary differential 
equation (\ref{eqte}) can be solved for $f(t)$ by standard means, e.g. by the
method of variation of constants. Introducing $f(t)=c(t) \exp(bt)$ and solving
the resulting equation $\partial_t c=\exp(-bt)\partial_t \overline H$  
yields 
\begin{equation}
f(t)={\rm e}^{bt}\left[c(0)+\int_0^t dt^\prime\,{\rm e}^{-bt^\prime}
\partial_{t^\prime} 
\overline H(t^\prime)\right].
\label{eqtf}
\end{equation}
In general, the integration constant $c(0)$ in (\ref{eqtf}) can be arbitrarily
chosen. This is a consequence of the afore-mentioned invariance of 
Eq.(\ref{eqtb}) with respect to $\hat h\rightarrow \hat h+\exp(bt)z$ 
($z$ arbitrary and constant) which implies that there is actually a continuous 
number of transformations leading from Eq.(\ref{eqta}) to Eq.(\ref{eqtb}).
A convenient choice, however, is to select as initial condition for 
$\hat h$ that $h({\bf x},t)$ and $\hat h({\bf x},t)$ initially coincide, 
i.e. $h({\bf x},0)=\hat h({\bf x},0)$. 
Consequently, $c(0)=-F_0/b$ holds and the transformation reads     
\begin{equation}
h({\bf x},t)=\hat h({\bf x},t)-\frac{F_0}{b}(1- {\rm e}^{bt})-{\rm
e}^{bt}\int_0^t dt^\prime\,{\rm e}^{-bt^\prime}\partial_{t^\prime} 
\overline H(t^\prime). 
\label{eqtg}
\end{equation} 
The transformation (\ref{eqtg}) that reduces the nonlocal equation 
(\ref{eqta}) in $h$-system to a local equation (\ref{eqtb}) in 
$\hat h$-system has several specific properties. (i) It is a purely 
{\it temporal} (integral) transformation relating $h({\bf x},t)$ 
in the physically
meaningful coordinate system comoving to the mean evolution of the 
erosion depth with the evolution of $\hat h({\bf x},t)$ in a temporally 
shifted system with no obvious physical significance. This shift 
$s(t)=h({\bf x},t)-\hat h({\bf x},t)$ being 
represented by the last two terms in (\ref{eqtg}), takes over the nonlocal 
properties of Eq.(\ref{eqta}) and is therefore nonlocal by virtue of the last 
term in (\ref{eqtg}). Noteworthy, $s(t)$ integrates over the temporal 
history of the mean velocity of the erosion front via $\partial_{t^\prime} 
\overline H(t^\prime)$. 
(ii) The stochastic part in Eq.(\ref{eqta}) remains unchanged by the 
transformation.  
(iii) The transformation (\ref{eqtg}) is highly 
useful because it allows for the direct applicability of theoretical and
numerical results obtained from Eq.(\ref{eqtb}) to Eq.(\ref{eqta}). 
Note, however, 
that for a correct interpretation in the physical meaningful $H$- or $h$-system 
the full temporal information of the mean evolution of the erosion front 
$\overline H(t)$ needs to be separately determined. If the dynamics of
$\hat h({\bf x},t)$ is known, this can be achieved by using
\begin{equation}
\overline H(t)= F_0 t + \int_0^t dt^\prime \,\left[\overline{G[\nabla \hat h)]}
+\overline\eta(t^\prime)\right].
\label{trafo11}
\end{equation}
 (iv) The basic prerequisite for the transformation (\ref{eqtg}) is that 
Eq.(\ref{eqta}) is linear in $h$. Extensions to nonlinear dependences on $h$ 
in Eq.(\ref{eqta}) do not seem to be generally feasible.    

For the specific case under consideration, 
$G[\nabla h]=a_1 \nabla^2 h+a_2\nabla^4 h+a_3(\nabla h)^2$, the transformation 
(\ref{eqtg}) can be further simplified. Using (\ref{trafo3}) it follows that
\begin{equation}
h({\bf x},t)=\hat h({\bf x},t)
        -a_3\int_0^t dt^\prime\,{\rm e}^{-b(t^\prime-t)}
	  \overline{(\nabla h)^2}(t^\prime)
\label{eqtk}
\end{equation}
and, consequently,
\begin{equation}
\partial_t \hat h= b \hat h+a_1 \nabla^2 \hat h
       +a_2\nabla^4 \hat h+a_3(\nabla \hat h)^2+\eta
\label{eqtx}
\end{equation}
which constitutes the damped KS equation in the $\hat h$-system and possesses 
the obvious invariance under the transformation $\lbrace\hat h,a_3\rbrace
\rightarrow\lbrace -\hat h,-a_3\rbrace$.
Eq.(\ref{eqtk}) shows the importance and the genuine interrelation of 
the KS-type nonlinearity for the non-triviality of the transformation. 
If $a_3$ equals zero, then simply $h({\bf x},t)=\hat h({\bf x},t)$ follows. 

This argument can be generalized: Separating the functional 
$G[\nabla h]$ in Eq.(\ref{eqta}) in terms that can be rewritten as 
the divergence of a flux, 
$G_{\rm F}[\nabla h]=\nabla\cdot{\bf j}_{\rm F}$, and terms 
$G_{\rm NF}[\nabla h]$ that cannot, the corresponding transformation 
is determined by  
$h({\bf x},t)=\hat h({\bf x},t)
        -\int_0^t dt^\prime\,{\rm e}^{-b(t^\prime-t)}
	  \overline{G_{\rm NF}[\nabla h]}(t^\prime)$. 
Consequently, any term being nonlinear and not originating 
from a flux leads to analogous complications as the term 
$\overline{(\nabla h)^2}$. Only if $G_{\rm NF}[\nabla h]=0$, then 
again  $h({\bf x},t)=\hat h({\bf x},t)$ holds implying 
that $\overline{H}(t)=F_0 t$.
\section{Some further results}
\subsection{Role of the nonlocal term} 
The general idea behind model equations 
such as Eq.(\ref{trafo2}) is that any individual term entering into the 
sum on the right hand side has its individual physical significance. 
As argued by Facsko et al. \cite{facsko-t}, the nonlocal term in 
(\ref{trafo2}) might be 
interpreted as a redepostion effect of eroded particles. To clarify the role 
of the nonlocal term in Eq.(\ref{trafo2}), $b (H-\overline H)$, 
we disregard, for the moment, all terms in Eq.(\ref{trafo2}) that 
depend on derivatives of $H$ and the stochastic fluctuations $\eta$. 
So, the equation
$\partial_t H=F_0 +b (H-\overline H)$
remains to be solved. 
Obviously $\partial_t\overline H=F_0$ holds and, consequently, (i) the 
mean depth evolution increases linearly with time for all times $t$ 
according to $\overline H=F_0 t$ and (ii)  $H-\overline H$ can 
be rewritten as $H-F_0 t$ showing that this term, individually considered, 
 acts locally in the frame moving with $F_0 t$. 
The solution for the erosion profile $H$ 
is then simply given by $H({\bf x},t)=H({\bf x},0)\exp(bt)+F_0 t$. 
For negative $b$, the impact of the term $b (H-\overline H)$ is just 
an exponential diminishment of the initial target profile 
$H({\bf x},0)$ in such a way that the overall shape and, in particular, 
the maxima, minima, and inclination points of $H({\bf x},t)-F_0 t$ remain 
at same spatial positions. Only the amplitude of $H({\bf x},t)$ decays in time. 
Consequently, it is not obvious whether the term $b (H-\overline H)$ 
can be interpreted as a redeposition effect that should also lead to a lateral 
variation of the erosion profile.     

\subsection{Stability of a flat erosion front}  
Taking advantage of the 
afore-mentioned arguments on the role of the 
nonlinearity in (\ref{trafo2}), ignoring the stochastic fluctuations 
for the moment and linearizing Eq.(\ref{trafo2}), i.e. omitting  
the term $(\nabla H)^2$, yields $\overline H(t)=F_0 t$. Therefore, 
a flat erosion front given by $H_{\rm FF}({\bf x},t)=F_0 t$ or, 
equivalently $h_{\rm FF}({\bf x},t)=0$ solves the 
deterministic limit of the linearized version of Eq.(\ref{trafo2}) 
and determines its 
basic solution. Consequently, 
as far as the linear stability  of the flat erosion front solution is
concerned, the nonlocal term can be replaced by the local term $F_0 t$ 
in the linearized Eq.(\ref{trafo2}). Then, standard techniques can 
be invoked for the stability analysis of $H_{\rm FF}$: 
Using an ansatz for a perturbation of $h_{\rm FF}({\bf x},t)=0$ of the form  
$h\propto \exp[i {\bf k}\cdot{\bf x}+\sigma t]$, yields a dispersion relation
$\sigma=b-a_1 k^2+a_2 k^4$ ($k^2={\bf k}^2$) for the growth rate $\sigma(k)$ 
of perturbations with a wave vector ${\bf k}$. 
Therefore, $\sigma(k)$ has its maximum at 
$k^2_{\rm max}=a_1/2a_2$ with $\sigma(k_{\rm max})=b-a_1^2/4a_2$ and the flat
erosion front $H_{\rm FF}$ is stable if $b\leq a_1^2/4 a_2$.

\subsection{Time dependence of the mean erosion depth $\overline H$}   
A crucial point of our discussion is the fact that variations 
of the mean erosion velocity about $F_0$, 
$\partial_t\overline H-F_0=a_3\overline{(\nabla H)^2} +\eta$,  
are generally non-zero and not even constant during the course of the 
erosion process. Here, we substantiate this by numerical simulations of 
Eq.(\ref{trafo2}) for a representative set of parameter values that leads to 
hexagonal patterns and draw some further conclusions.

In Fig. 1 we show the evolution of the ensemble or stochastically 
averaged evolution of $\partial_t\overline H-F_0$ given by 
$\partial_t\langle\overline H\rangle-F_0=
a_3\langle\overline{(\nabla H)^2}\rangle$ since 
$\langle\overline\eta\rangle=\overline{\langle\eta\rangle}=0$. For this 
quantity, the stochasticity necessarily present in individual runs of the 
erosion process, as shown in Fig. 2, is leveled out. In both cases, the 
initial target profile has been kept fixed to a small Gaussian 
distribution of the initial amplitudes with a maximum amplitude 
of $\eta=0.01$ about the perfectly flat state.

The generic behavior as function of time consists of three parts: 
(i) For very small erosion times, 
$\partial_t\langle\overline H\rangle-F_0$ is very close to zero since 
here the nonlinearity in the evolution equation can be neglected and, 
therefore mainly the linear evolution of $H$ contributes. Since 
$\partial_t\overline H-F_0$ is proportional to  $a_3\overline{(\nabla H)^2}$ 
this leads to a purely exponential increase for short times as depicted in 
the insets of Fig. 1. (ii) With increasing erosion time, the amplitude $H$ 
eventually reaches values where the nonlinear term in Eq.(\ref{trafo2}) becomes
comparable in size to linear terms. In this crossover time range, 
$\partial_t\langle\overline H\rangle-F_0$ increases very rapidly up 
to a point where the subsequent increase drastically slows down. (iii) 
For longer times, where the nonlinearity in Eq.(\ref{trafo2}) is 
fully developed, the KS-nonlinearity $(\nabla H)^2$ mainly reduces the 
further increase of $\partial_t\langle\overline H\rangle-F_0$ almost 
to a constant. Our numerical simulations that went very far into the full nonlinear
regime, however, do not indicate a full saturation, but a slight 
systematic increase with time that depends almost linearly on $t$ in the 
simulated time range. Similar behavior for individual erosion processes 
(albeit slightly modified by its intrinsic stochasticity) can also be read 
off from Fig. 2. An important consequence of this dynamical behavior of 
Eq.(\ref{trafo2}) is the fact that the underlying surface morphology 
$H({\bf x},t)$ does not saturate in a steady state unless 
$\partial_t\overline H-F_0$ fully saturates into a constant value. To 
substantiate this statement, we show in the insets of Fig. 2 for two 
representative erosion times the corresponding morphology behavior that 
clearly indicates a non-steady evolution. 

Finally, we like to draw attention to a challenging point for further 
experiments. Due to the non-monotonic behavior of 
$\partial_t\langle\overline H\rangle-F_0$ as function of the erosion time 
that is triggered by the KS-nonlinearity, a specific experimental 
measurement of this quantity can lead to experimental evidence for the 
existence and dominance of no-flux terms as discussed in section III.

\section{Summary}
To conclude, we stress the most important points of our investigation:
(i) The nonlocal, damped KS equation suggested by Facsko et al. 
\cite{facsko-t} can be obtained as minimal model from balance equation
considerations if nonlocal terms compatible with the underlying invariances are
allowed. Most importantly, such a nonlocal term being proportional to
$H-\overline{H}$ leads in combination with the KS nonlinearity to a
non-monotonic time evolution of the spatially averaged erosion depth 
and, connected with that, to a surface morphology that does 
not generally seem to approach a steady state (at least on the simulated 
time scales). (ii) The necessity for a non-local dependence arises from 
the fact that the KS nonlinearity cannot be rewritten in form of the divergence 
of a flux. If all terms entering into the evolution equation could be rewritten 
in form of such divergences, the non-local term could be replaced by a simple 
drift term $\overline{H}=F_0 t$. (iii) By using a specific transformation
presented in section III, the nonlocal, damped KS equation can be exactly 
recast in form of a standard (non-local) damped KS equation. This property 
greatly simplifies the analysis of the evolution equations. Finally, we have
also briefly proposed a simple experimental measurement in order to substantiate the 
pronounced effect of the KS nonlinearity or at least related non-flux
nonlinearities during the erosion process. Building on the results presented
here, we will discuss, in a subsequent publication, the rich pattern forming 
structure of an appropriately generalized non-local anisotropic damped 
KS equation for the case of oblique incidence.

\section*{Acknowledgments} 
We thank T. Allmers, M. Donath, and S. Facsko for interesting discussions.

\clearpage

\begin{figure}
	\centering
	
	  \includegraphics[width=.4\textwidth, clip, viewport=138 576 320 716]{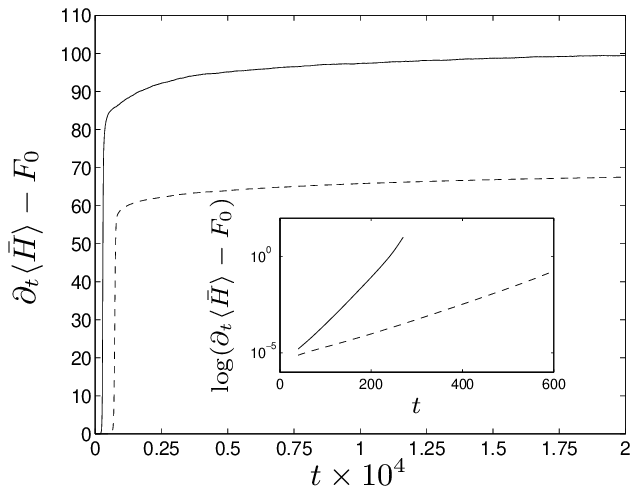}
		\caption{Growth rate of the mean erosion depth calculated 
		from Eq.(\ref{trafo2}), averaged over 100 runs. Parameters 
		are the same as in the paper by Facsko et al. 
		\protect{\cite{facsko-t}}, 
		except for a spatial step size of $dx =1$ 
		(mesh size $400\times 400$,$b=-0.24$, $a_1=-1$, $a_2=-1$, 
		$a_3=0.0025$, white noise with a maximum amplitude 
		of $\eta=0.01$). Solid line: $b=-0.22$, dotted line: 
		$b=-0.24$. The inset shows a semilogarithmic plot of a 
		portion of the data where linear terms determine 
		the growth. A linear fit in this regime yields 
		$\ln (\partial_t \langle\bar{H}\rangle- F_0)
		\propto 0.018\cdot t$ for $b=-0.24$ and 
		$\ln (\partial_t \langle\bar{H}\rangle- F_0)
		\propto 0.057\cdot t$ for $b=-0.22$. 
		This compares well to the result which can be 
		obtained from the linearized version of Eq. (\ref{trafo2}), 
		i.e. $2\sigma_{max}=0.02$ for $b=-0.24$ and 
		$2\sigma_{max}=0.06$ for $b=-0.22$.}
	\label{dthgm}
\end{figure}

\newpage

\begin{figure}
	\centering
	\includegraphics[width=.4\textwidth]{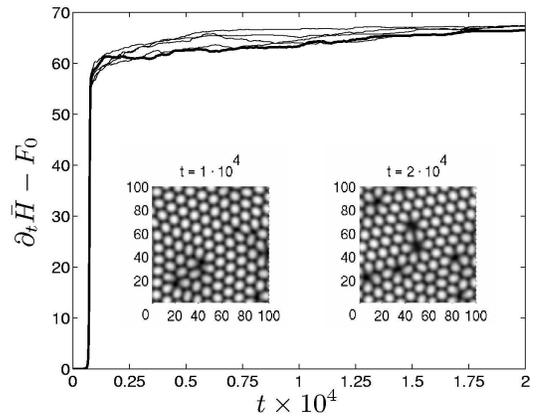}
	\caption{Growth rate of the mean erosion depth calculated 
	from Eq.(\ref{trafo2}) for five different noise sequences. 
	Parameters are the same as in Fig. \protect{\ref{dthgm}} 
	with $b=-0.24$. 
	Insets show a section of the surface morphology corresponding 
	to the thick line in the figure at times $t=1\cdot 10^4$ and 
	$t=2\cdot 10^4$.}	
\end{figure}

\end{document}